\documentclass[final,10pt,times,twocolumn]{elsarticle}

\usepackage{natbib}
\usepackage{graphics}
\usepackage{graphicx}
\usepackage{epsfig}

\usepackage{amssymb}

\journal{Electrochemistry Communications}

\begin{document}

\begin{frontmatter}

\title{Theory of Electric Double Layer Dynamics at Blocking Electrode}

\author{Maibam Birla Singh and Rama Kant*}

\address{Department of Chemistry,\\ 
University of Delhi,\\ 
Delhi 110007, India\\
Email address: rkant@chemistry.du.ac.in \\
URL: http://people.du.ac.in/$\sim$rkant}

\begin{abstract}

A phenomenological theory of electric double layer polarization of blocking electrode is presented solving modified Debye-Falkenhagen (MDF) equation for potential, under the impedance boundary condition. The dynamic impedance and capacitance are obtained in terms Debye screening length  and frequency dependent polarization length. Two characteristic relaxation frequencies for compact layer and diffuse layer are identified. At frequencies less than  Helmholtz layer relaxation frequency  the EDL is perfectly blocking.  At frequencies larger than the diffuse layer relaxation  frequency the EDL behaves like a resistor.  At crossover frequencies the electrode is polarized. Theoretical results agrees well with experimental capacitance  dispersion data.
\end{abstract}


\end{frontmatter}

\section{Introduction}

Electrode polarization \cite{Barsoukov, Schwan,Barbero,Ferry,Macdonald0, Buck,Gunning, Bazant1,Macdonald3} is a ubiquitous phenomenon and is found to have an increasing application in supercapacitors \cite{Mariappan}, batteries \cite{Park}, fuel cells \cite{Biesheuvel09}, solar cells \cite{Bisquert2000} and  ion desalination membrane electrodes \cite{Biesheuvel}. Relaxation and dispersion \cite{Serghei,Amstrong1, Amstrong2} are two electrochemical phenomena usually observed when the electrode is coupled to various electrolytes viz., aqueous, liquid crystals \cite{Barbero}, glassy electrolytes \cite{Mariappan}, polyelectrolytes \cite{Klein} and ionic liquids \cite{Serghei}. 

In this communication, we report a model of electrode polarization in EDL formed near a planar blocking electrode (Fig.1). The EDL is divided into two region, compact layer and diffuse layer separated by outer Helmholtz plane (OHP). The objective of the paper is to understand the interplay between the compact and diffuse layers in-terms of the length scale and various time scales emerging out of the electric double layer phenomena when a  sinusoidal time varying potential is applied.

\section{Model and Solution} 
The Debye-Falkenhagen equation \cite{Buck, Bazant1, Falkenhagen} for relaxation of the ionic atmosphere in diffuse layer at the planar electrode  is  
\begin{equation}
\frac{1}{D} \frac{\partial \rho}{\partial t} = (\nabla^2  - \kappa^2) \rho \label{E6}
\end{equation}
where $D$ is the diffusion coefficient, $\rho = e (z_{+} c_{+} - z_{-} c_{-})$ is the charge density and $c_{\pm}$ is th ionic concentrations, $z_{\pm}$ the ionic charges and $\kappa^{-1}$ is the Debye screening length
\begin{equation}
 \kappa^{-1} = \sqrt{\frac{\epsilon_{r} \epsilon_{0} k_{B} T} {2 N_{A} e^2 I}}  
\end{equation}
where $I = (1/2)\sum_{i=1}^{n} c_{i} z_{i}^2$ is the ionic strength, $k_B$ the Boltzmann's constant, $N_{A}$ is tha Avogadro number,  $T$  the temperature,  $\epsilon_{0}$ the  permittivity of free space and $\epsilon_{r}$ dielectric constant of the solvent. Now we assume near equilibrium local concentration ions $c_{i}$ is  given by Boltzmann equation 
\begin{equation}
c_{i} = c_{i} ^{\infty} \exp (\pm z_{i} e \phi /k_B T) \label{E7}
\end{equation}
Now for a symmetrical $ z_{+} = z_{-} = z $ electrolytes and assuming $ c_{+} ^{\infty} = c_{-} ^{\infty} = c_{0}$, linearizing Eq. (\ref{E7}) under condition $ \phi \ll k_{B} T/z e$ and substituting in Eq. (\ref{E6}), we obtained the modified Debye-Falkenhagen (MDF) equation for potential $\phi$ 
\begin{equation}
\frac{1}{D} \frac{\partial \phi}{\partial t} = (\nabla^2  - \kappa^2) \phi  \label{E8}
\end{equation}
For a sinusoidal applied potential, $\phi = \phi(r) e^{i \omega t } $ the MDF equation is 
\begin{equation}
\left [\nabla^2  - \kappa^2\, ( 1+ i\, \omega \tau_{D})\right ] \phi (r) = 0 \label{E9}
\end{equation}
where  $\tau_{D} = 1/\kappa^2 D$ is the Debye time \cite{Macdonald}. Eq.~(\ref{E9}) represent the dynamic picture of the interface, when a sinusoidal potential is applied at the interface (Fig.~1). 

 \begin{figure}[htbp]
\vskip+0.15in
\includegraphics[scale = 0.6]{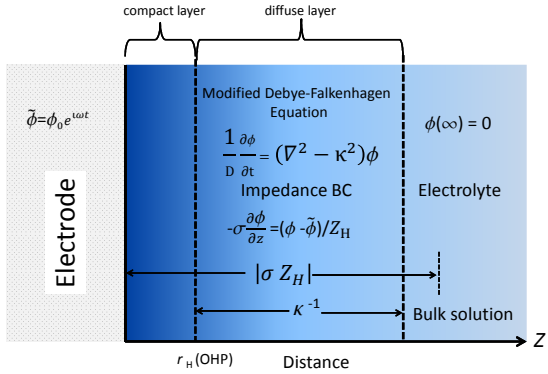}
\caption{Schematic diagram of electric double layer (EDL) formed at planar blocking electrode. }
\end{figure}

In the case of blocking electrode, the ohmic current ($j$) crossing a metal/electrolyte interface is given by
\begin{equation}
j = - \sigma \nabla \phi \label{E12}
\end{equation}
where $\sigma$ is specific interfacial conductivity of electrolyte near electrode.
The interfacial and bulk electrolyte conductivity ( $\sigma_0$) satisfies $\sigma_{0} \gg \sigma$ \cite{Serghei}, where $\sigma_0= \epsilon_0 \epsilon _r \kappa D$ \cite{Barbero,Bazant1}.
 Eq.~(\ref{E12}) gives the current arriving at the OHP and is equal to the ratio of potential drop at the OHP ($\phi- \tilde \phi $), to the impedance  ($Z_H$) of the Helmholtz layer, i.e. $(\phi - \tilde \phi)/ Z_H$  where $\tilde \phi$ is the potential of the electrode and $\phi$ is potential at the OHP. Now equating the current on both sides  we have the impedance boundary condition \cite{Ball}
\begin{equation}
-\sigma (\partial \phi/\partial z) = (\phi -\tilde \phi)/Z_{H} \label{E15}
\end{equation}
where $Z_{H} = ({1}/{i \omega C_H +R_{H}})$ is the impedance of the Helmholtz layer and $R_H$  and $C_H$ 
is the compact layer resistance and capacitance. 
The other semi-infinite (bulk) boundary condition is $ \phi (\infty) = 0$.                                                                                                                                                                                                                                                                                                    
The impedance boundary condition may also be written as $\phi = \tilde \phi -\Lambda_{C}(\omega)^{-1}
(\partial \phi/\partial z)$ where $\Lambda_{C}(\omega) = \sigma Z_{H}$. The quantity $\mid\Lambda_{C}
(\omega)\mid ^{-1}$ is the phenomenological frequency dependent length. It is the effective thickness of interfacial  layer where the electrode is polarized and decides the crossover time scales from purely capacitive 
regimes to resistive regimes we called it the ``polarization length". It may be rewritten for simplicity as $\Lambda_{C}(\omega) =  1/ (\sigma/i\, \omega C_H + W)$ where $W = \sigma R_{H}$ is a characteristic resistive length which may be called as modified Wagner number \cite{Bosco, Wagner}. The value of $\mid\Lambda_{C}(\omega)\mid^{-1}$ is decided by the compact layer capacitance $C_{H}$, resistance $R_{H}$ and the conductivity of the electrolyte $\sigma$.

We use the method developed in ref.~\cite{Kant} to obtained solution of MDF Eq.~(\ref{E9}). The  total admittance of  planar blocking electrode is
\begin{equation}
Y_{P}(\omega)  =  \frac{A_{0}} {Z_{H} +1/(\sigma \kappa ( 1+ i \omega \tau_{D})^{1/2})} \label{E19}
\end{equation}
where A$_0$ is the area of electrode. In order to understand behavior of admittance, we consider the following frequency regimes. \\

At low frequency $\kappa^{-1} \ll \mid \Lambda_{C}(\omega)\mid^{-1} $, from Eq.~(\ref{E19}) we have 
\begin{equation}
Y_{P}(\omega) =  \left(\frac{A_0}{Z_{H}}\right) \left[1- \frac{\Lambda_{C}(\omega)}{\kappa} + \left(\frac{\Lambda_{C}(\omega)}{\kappa}\right)^2 +\cdots \right] \label{E22}
\end{equation}
The leading contribution in admittance originates from the Helmholtz layer and its characteristic relaxation frequency is $\omega_{H} = (C_H R_H)^{-1}$.
 When the frequency is $\omega \ll \omega_{H}$, Eq.~(\ref{E22}) shows that the EDL behaves like an ideally polarizable blocking electrode. \\

At high frequency $\mid\Lambda_{C}(\omega)\mid^{-1} \ll \kappa^{-1} $, from Eq. (\ref{E19}) we have
\begin{equation}
Y_{P}(\omega) = A_0 \sigma \kappa \left[1- \frac{\kappa}{\Lambda_{C}(\omega)} + \left(\frac{\kappa}{\Lambda_{C}(\omega)}\right)^2 +\cdots \right] \label{E23}
\end{equation}
When the frequency is higher than the characteristic diffuse layer relaxation frequency, $\omega_{D}< \omega$, 
Eq.~(\ref{E23}) shows that the leading order contribution to the interfacial admittance comes from the diffuse layer resistance. 
This observation will allow us to see the effects of concentration on the admittance of planar electrode. The admittance is resistive and the characteristic frequency at which the EDL  relaxes is the characteristic frequency at which diffuse layer relaxes, $\omega_{D} = (\sigma \kappa)/C_{D} = \sigma/  (\epsilon_0\epsilon_{D})$. But $ \sigma/(\epsilon_0 \epsilon_{D})$, is also a characteristic  conductivity relaxation frequency \cite{Barsoukov}.

At the crossover region, the frequency range is confined to $ \omega_{H}< \omega< \omega_{D} $ and  $\mid\Lambda_{C}(\omega)\mid^{-1}$ is comparable to $\kappa^{-1}$. In this case the crossover frequency is resultant of both processes- relaxation of Helmholtz layer and the relaxation of diffuse layer; and we expect the dispersions in electrochemical response of EDL. 

Now expanding Eq.~(\ref{E19}) in the ratio of the Debye screening length to ion diffusion length, we have
\begin{eqnarray}
 Y_{P}(\omega) & = &  A_0 \kappa \sigma \left[ \frac{1}{1+ \kappa\sigma Z_{H}} +  \frac{i }{2(1+ \kappa\sigma Z_{H})^{2}} \left(\frac{\omega}{\kappa^2 D}\right)^{2} \right.
\nonumber\\
&&\left.+  \frac{1+3 \kappa \sigma Z_{H}}{8(1+ \kappa\sigma Z_{H})^{3}}  \left(\frac{\omega}{\kappa^2 D}\right)^{4} - \cdots \right]. 
\end{eqnarray} 
At high frequencies,  $\kappa^{-1}/\sqrt{D/\omega}> 1$, then the higher terms will contribute. Gunning et al  \cite{Gunning} obtained the impedance depended on  zeta potential of the diffuse double layer. The theory developed here differs in the inclusion of the \emph{Stern} layer along with the diffuse double layer. Our theory  focuses on the boundary constraint at the OHP rather than the ``slipping plane" \cite{Gunning}.

\section{Results and Discussions} 
\begin{figure}
\vskip-0.1in
\includegraphics[scale = 0.6]{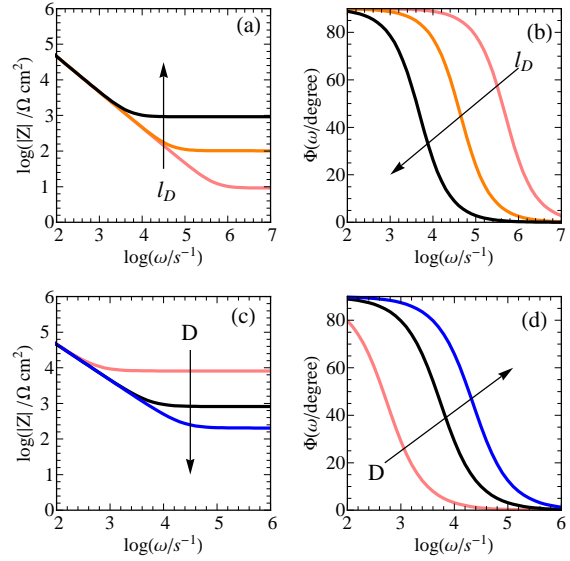}
\caption{ Effect of concentration variation is shown in  (a) and (b). The Debye length $\kappa^{-1}$ is varied from 0.3, 1 and 3. Effect of electrolyte conductivity variation  is shown in (c) and (d). The diffusion coefficient $D~ (10^{-5} cm^2/s$) is varied from 0.02, 1 and 4. The graphs are plotted using  $\epsilon_0 $ = 8.854 $\times 10^{-14}$ (F/cm),  $\epsilon_H$ = 3.5, $r_0 $= 0.53 nm and area of electrode $A = 0.038 cm^2$  }
\end{figure}

Fig. 2 (a) show the effect of concentration on the log-log plot of magnitude of  impedance, $|Z(\omega)|$ and frequency $\omega$. As we increase the concentration of electrolyte  the  crossover frequency (system changes    from capacitive to resistive behavior) decreases. The frequency dependence of electrode polarization is shifted to  lower frequencies as the concentration of the electrolyte decrease. 
Fig. 2 (b) shows the phase angle  $\Phi$ vs log~$\omega$. The plot shows that the impedance behavior is resistive at the high frequency, which is identified in the plot as $\Phi \rightarrow 0^{\circ}$.  The capacitive behavior is seen at a low frequency which is identified in the plot as $\Phi \rightarrow 90^{\circ}$. As we decrease the concentration of electrolyte the crossover frequency at which the  phase behavior  switch from capacitive to resistive behavior is shifted progressively to lower frequencies. Fig. 2 (c) show the influence of electrolyte conductivity through the diffusion coefficient of ions on the log-log plot of magnitude of  impedance, $|Z(\omega)|$ and frequency $\omega$. With the increase in the diffusion coefficient of electrolyte  the frequency at which the system changes from capacitive to resistive behavior increases. The frequency dependence of electrode polarization occurs at a higher frequency as we increase the conductivity (by increasing the diffusion coefficient). Thus in viscous solutions where the diffusion coefficient of ion is low the frequency dependent behavior is shifted progressively to lower frequencies. Fig. 2 (d) shows the phase angle  $\Phi$ vs log$~\omega$. The plot shows that the impedance behavior is resistive at the high frequency, which is identified in the plot as $\Phi \rightarrow 0^{\circ}$.  The capacitive behavior is seen at a low frequency which is identified in the plot as $\Phi \rightarrow 90^{\circ}$. As we increase conductivity of electrolyte  the characteristic frequency at which the  phase behavior  crossover from capacitive to resistive behavior is shifted progressively to higher frequencies.

Fig. 3 shows the theoretical plots fits with experimental capacitance dispersion data for Hg/electrolyte interface. The real part of the capacitance is calculated as $C'(\omega) =  Z '' (\omega) / \omega |Z (\omega)|^{2}$.
In agreement with the theory, the data shows two crossover frequencies. The effect of electrolyte conductivity on capacitance is exactly same as in experiment. Increasing the electrolyte conductivity by increasing diffusion coefficient, increases the crossover frequency $(\omega_{D}$) of capacitance dispersion.
\begin{figure}[htpb]
\includegraphics[scale = 0.55]{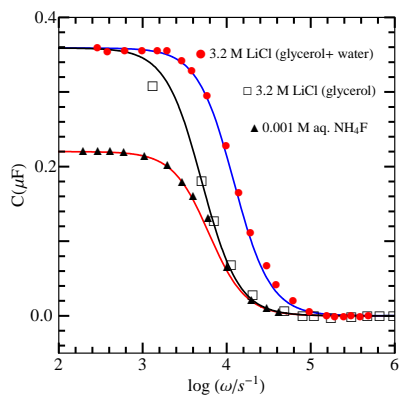}
\caption{Comparison of theory with experiment. Data were taken from ref. \cite{Amstrong1, Amstrong2}.  Physical parameters used to fit the data are listed in table 1.}
\end{figure}
 
\section{Conclusions} 

The electrode polarization is strongly dependent on the phenomenological polarization length.  The Helmholtz and diffused layers relaxation frequencies  divide the whole frequency range into three regimes. The low frequency regime is characterized by  large polarization layer thickness and the capacitance of EDL is that of  Helmholtz layer. Whereas the high frequency is characterized by small polarization layer and the capacitance of the EDL is that of diffuse layer. The crossover frequency is indicated by a transition from capacitance to resistive behavior and  an S-shaped crossover region in the capacitance plot. The capacitance dispersion that arises in the crossover  regime is found to depend on the concentration and conductivity of the electrolyte. Decrease in the concentration of electrolyte results in the shift of the crossover region progressively to lower frequencies. The capacitance dispersion is seen for frequencies where $\omega_H < \omega < \omega_D$.  The  theory agrees  with the capacitance dispersion data of the mercury/electrolyte interface for an aqueous, glycerol and glycerol-water system. 
Finally one can say, this theory is an indispensable step in the quantitative description of the dynamics of planar blocking electrode. 

\begin{small}
\begin{table}[t]
\small\addtolength{\tabcolsep}{-3pt}
\caption{Here is $\epsilon_{H}$  dielectric constant and $r_{H}$ (nm) is effective thickness (corresponding to hydrated ion size) of Helmholtz layer. The effective diffusion coefficient $D~ (10^{-6} cm^2/sec)$ used as fitting values are listed. Here we have assumed that two conductivities follow: $\sigma = k\, \sigma_0$ \cite{Serghei} and $k = 10^{-6}$.}
\begin{tabular}{  r    r    r    r    r    r    r    r    r   r}
\hline   
Data & E  & S & $ C_{b} $ &  $\epsilon_{H} $ &  $\epsilon_{G}$ &  $r_{H}$ &  A$_0$ & $D $\\
\hline
Fig.5 (a)     &   NH$_4$F & H$_2$O &   10$^{-3}$  & 3.5 & 78.6  &  0.53$^{a}$ &38 & 18.3  \\
Fig.5 (b)     &   LiCl & gly &   3.2  & 6 & 42.5$^{b}$ & 0.37 $^{a}$ & 25 & 0.06 \\
Fig.5 (b)     &   LiCl & gly+H$_2$O &  3.2  & 6 & 42.5$^{b}$ & 0.37$^{a}$ & 25 & 15 \\
\hline
\end{tabular}
\newline
\intextsep0.1in {\footnotesize{E- Electrolyte, S- solvent,gly- Glycerol, $C_{b}$- molar concentration,  $A_{0}$ (10$^{-3}$)- area of electrode in cm$^{2}$ , a- \cite{Nightingale},   b- \cite{Robinson}}}
\noindent
\end{table}
\end{small}

\end{document}